\documentclass[apsf,pra,showpacs,amsmath,amssymb,twocolumn,nofootinbib,superscriptaddress,longbibliography]{revtex4-2}

\usepackage[pdftex]{graphicx}

\usepackage{mathrsfs,amsmath,amssymb,amsthm,graphicx}
\usepackage{fancyhdr}
\usepackage{caption}
\usepackage{subcaption}

\usepackage{braket}
\usepackage[colorlinks=true,
            linkcolor=red,
            urlcolor=blue,
            citecolor=blue]{hyperref}
 
\draft % marks overfull lines with a black rule on the right

\begin{document}
\bibliographystyle{apsrev}

\title{Two-Way Quantum Time Transfer: A Method for Daytime Space-Earth Links}
\author{Randy Lafler}
\affiliation{Air Force Research Laboratory, Directed Energy Directorate, Kirtland AFB, NM, United States}

\email[AFRL.RDSS.OrgMailbox@us.af.mil\\
DISTRIBUTION A: Approved for public release; distribution is unlimited. \\
Public Affairs release approval AFRL-2024-0986]{}

\author{Mark L. Eickhoff}
\affiliation{The Boeing Company, Albuquerque NM, United States}

\author{Scott C. Newey}
\affiliation{The Boeing Company, Albuquerque NM, United States}

\author{Yamil Nieves Gonzalez}
\affiliation{The Boeing Company, Albuquerque NM, United States}

\author{Kurt E. Stoltenberg}
\affiliation{The Boeing Company, Albuquerque NM, United States}

\author{J. Frank Camacho}
\affiliation{Leidos, Albuquerque NM, United States}

\author{Mark A. Harris}
\affiliation{Leidos, Albuquerque NM, United States}

\author{Denis W. Oesch}
\affiliation{Leidos, Albuquerque NM, United States}

\author{Adrian J. Lewis}
\affiliation{Air Force Research Laboratory, Directed Energy Directorate, Kirtland AFB, NM, United States}

\author{R. Nicholas Lanning}
\affiliation{Air Force Research Laboratory, Directed Energy Directorate, Kirtland AFB, NM, United States}

\date{\today}

\begin{abstract}
High-precision remote clock synchronization is crucial for many classical and quantum network applications. 
Evaluating options for space-Earth links, we find that traditional solutions may not produce the desired synchronization for low Earth orbits and unnecessarily complicate quantum-networking architectures.
Demonstrating an alternative, we use commercial off-the-shelf quantum-photon sources and detection equipment to synchronize two remote clocks across our freespace testbed utilizing a method called two-way quantum time transfer (QTT).
We reach picosecond-scale timing precision under very lossy and noisy channel conditions representative of daytime space-Earth links and software-emulated satellite motion.
This work demonstrates how QTT is potentially relevant for daytime space-Earth quantum networking and/or providing high-precision timing in GPS-denied environments.
\end{abstract}

\maketitle %\maketitle must follow title, authors, abstract
\pagestyle{fancy}
\cfoot{DISTRIBUTION A: Approved for public release; distribution is unlimited. Public Affairs release approval AFRL-2024-0986.}
\lhead{}
\chead{}
\rhead{\thepage}

%\section{Introduction} 
\label{sec: Introduction}
Precise synchronization of remote clocks is important for position, navigation, and timing (PNT), high speed transactions, distributed computing, quantum networking, and many other applications.
One can synchronize remote clocks with global positioning system (GPS) public signals and achieve nanosecond-scale synchronization\cite{lombardi2001time}.
If more precision is desired, or one is operating in a GPS-denied environment, other techniques must be used.
Perhaps the most straightforward optical-time-transfer technique uses pulsed lasers, photodetectors, and software-based correlation methods.
For example, time transfer by laser link (T2L2) demonstrations have achieved picosecond-scale precision between remote ground stations operating in common view with the Jason-2 satellite after a 1000-s acquisition \cite{exertier2014time}.
Optical two-way time and frequency transfer (O-TWTFT) utilizes frequency combs to synchronize two remote clocks to femtosecond precision \cite{giorgetta2013optical,sinclair2016synchronization}.  
To date, demonstrations of O-TWTFT have been performed between stationary sites \cite{giorgetta2013optical,sinclair2016synchronization,caldwell2023quantum} and slow-moving drones with velocities $<25$ m/s, but there are anticipated problems with the large Doppler shifts associated with low Earth orbits (LEO) that have not been resolved \cite{caldwell2023quantum}.
Furthermore, femtosecond-scale synchronization based on O-TWTFT may be excessive for some applications and thus not justify the additional cost and complexity.
Alternatively, the White Rabbit protocol can achieve picosecond-level precision and has been investigated for quantum-networking synchronization \cite{burenkov2023synchronization, alshowkan2022advanced, schatz2023practical}. 
It was designed for wireline fiber-optic applications but has been implemented wirelessly \cite{gilligan2020white} and considered for space applications \cite{jamrozy2015white}. 
However, these techniques all introduce systems and hardware that may unnecessarily complicate quantum-networking architectures.

A more direct and economical solution would utilize the femtosecond-scale temporal correlations of photon pairs that are already being exchanged for quantum-communication tasks.
In this case, the relative time offset between two remote clocks is measured with the following procedure: 
(1)~a series of photon pairs are separated and transmitted to two remote sites,
(2)~the photons are detected and their arrival times are time-tagged based on the respective local clock, 
(3)~after sufficient detection events are collected, the series of arrival times from each site are combined and correlation methods are used to find the clock offset.  
This technique was first proposed in Ref.~\cite{valencia2004distant} and we refer to it as quantum time transfer (QTT). 
One-way QTT enables relative clock synchronization \cite{ho2009clock}, and two-way QTT enables absolute clock synchronization \cite{lee2019symmetrical}.
Subsequently, we investigated the suitability of this method for lossy and noisy channel conditions commensurate with daytime Earth-satellite quantum downlinks \cite{lafler2023quantum}.
Others have emulated freespace links in a lab setting \cite{spiess2023clock} and investigated effects in global-scale architectures \cite{haldar2023towards, haldar2023synch}.

In this letter, we report remote clock synchronization and ranging with two-way QTT during atmospheric conditions representative of bi-directional Earth-satellite links and software-emulated satellite motion to $\mathcal{O}(1/c)$, that is, first-order Doppler effect \cite{blanchet2001relativistic}.
Our system achieves picosecond-scale timing precision during these challenging conditions with commercial off-the-shelf (COTS) hardware.
We discuss the measured clock offset, the overlapping Allan deviation, the time deviation, the emulated satellite range, and the coincidence-to-accidental ratio (CAR) of the QTT correlation signals relative to the sky noise.

\begin{figure}[t!]
	\includegraphics[width=0.85\columnwidth]{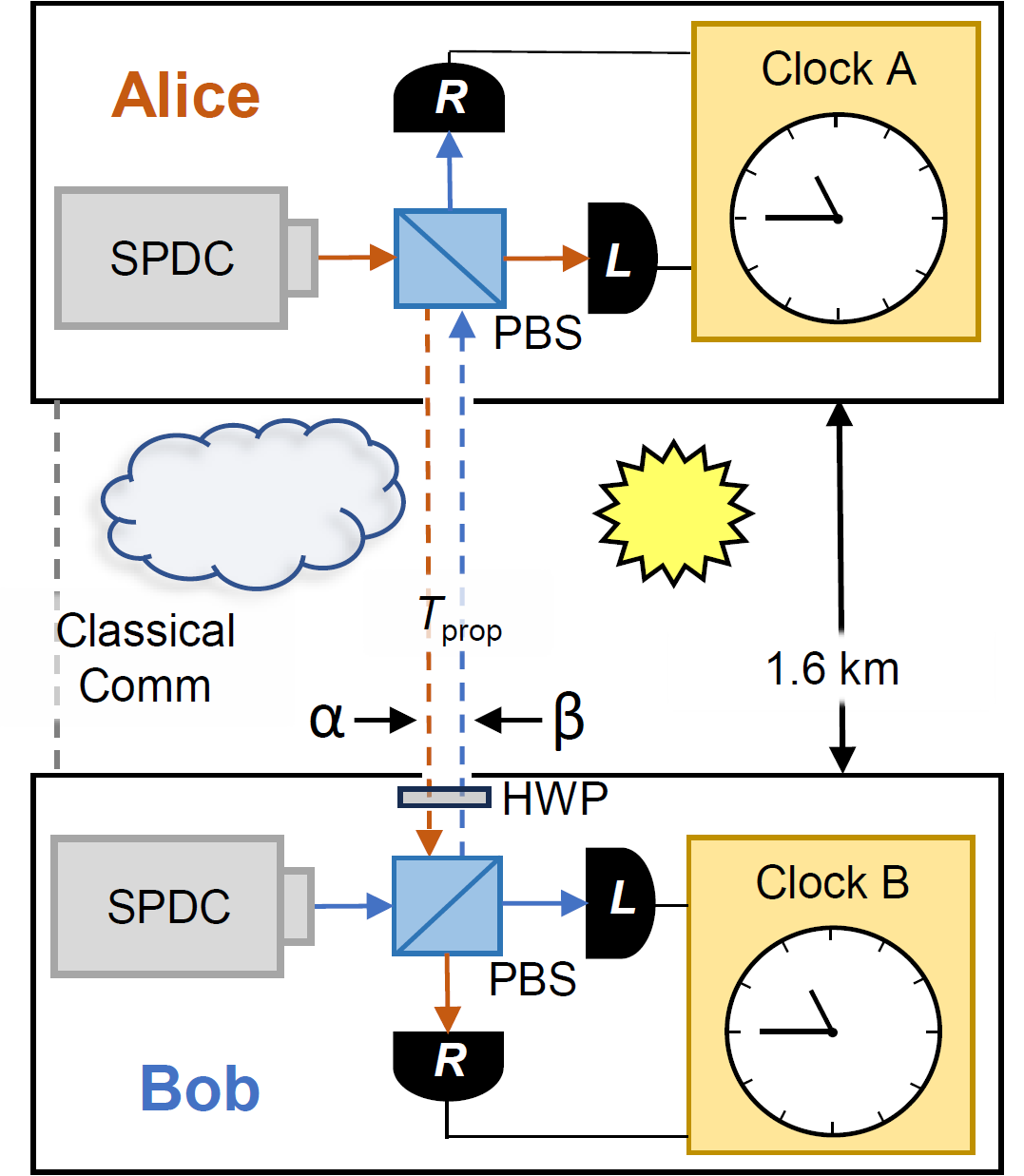}	
	\caption{
		Schematic of two-way QTT field experiment.  
		Each site is equipped with an SPDC bi-photon source, local $L$ and receive $R$ detectors, and a local clock consisting of a time tagger, a rubidium frequency standard, and a PC.  
		The sites are separated by a 1.6 km freespace channel
		with propagation time $T_{\mathrm{prop}}$.  
		The ``downlink'' direction $\alpha$ and ``uplink'' direction $\beta$ represent the directions for the emulated satellite pass.
		The classical channel is utilized for sharing time tags.
	}
	\label{fig:schematic}
\end{figure}

%\section{Experiment}
\label{sec: Field Exp}
Our freespace quantum-communication testbed is located at the Starfire Optical Range (SOR), Kirtland AFB, NM in the Southwestern United States.
The transceiver sites are the same sites utilized in Ref.~\cite{gruneisen2021adaptive}, but are enhanced for two-way propagation using the arrangement in Fig.~\ref{fig:schematic}.
The transceivers Alice and Bob represent satellite and ground station, respectively.
We tune the Alice-to-Bob direction to be representative of a low-earth orbit (LEO) downlink; we set the beam divergence to impose the correct geometric loss, we use a white light source to scatter noise photons into the channel, we use a heat source to add scintillation, and we utilize a closed-loop adaptive-optics (AO) system to monitor and compensate for atmospheric turbulence \cite{gruneisen2021adaptive}.

In the Bob-to-Alice ``uplink'' direction, the quantum signal was precompensated by the AO system, and photons arrived at Alice with slightly  better efficiency than the downlink direction due to less divergence.
For actual Earth-satellite uplinks, the quantum signal is transmitted at a point-ahead angle so that it intercepts the LEO satellite.
As a consequence, the uplink quantum signal travels a different path than the downlink quantum signal.
Depending on the path of the AO beacon, this can result in additional loss due to reduced compensation of atmospheric-turbulence effects caused by anisoplanatism \cite{tyson2015principles,  martinez2023atmospheric}. 
We apply an additional $2$ dB of attenuation to the true coincidences in the receive channel at the Alice site in order to make our ``uplink" direction attenuation the same as the ``downlink" direction.
This level of attenuation corresponds to AO compensation with an ideal beacon in the point-ahead direction \cite{oliker2019much}.
Performing the uplink with the AO loop closed on a \textit{downlink} beacon would result in $\sim$10 dB more attenuation \cite{oliker2019much}, but when applying this level of attenuation we observe occasional loss of synchronization due to the relatively low-performance sources in this field experiment.
In a practical quantum network, one would use a higher pair rate or heralding efficiency source at the ground station, which would increase the attenuation tolerance and relax the requirement for the beacon in the point-ahead direction \cite{lafler2023quantum}.

\begin{figure*}[t!]
	\includegraphics[width=1\linewidth]{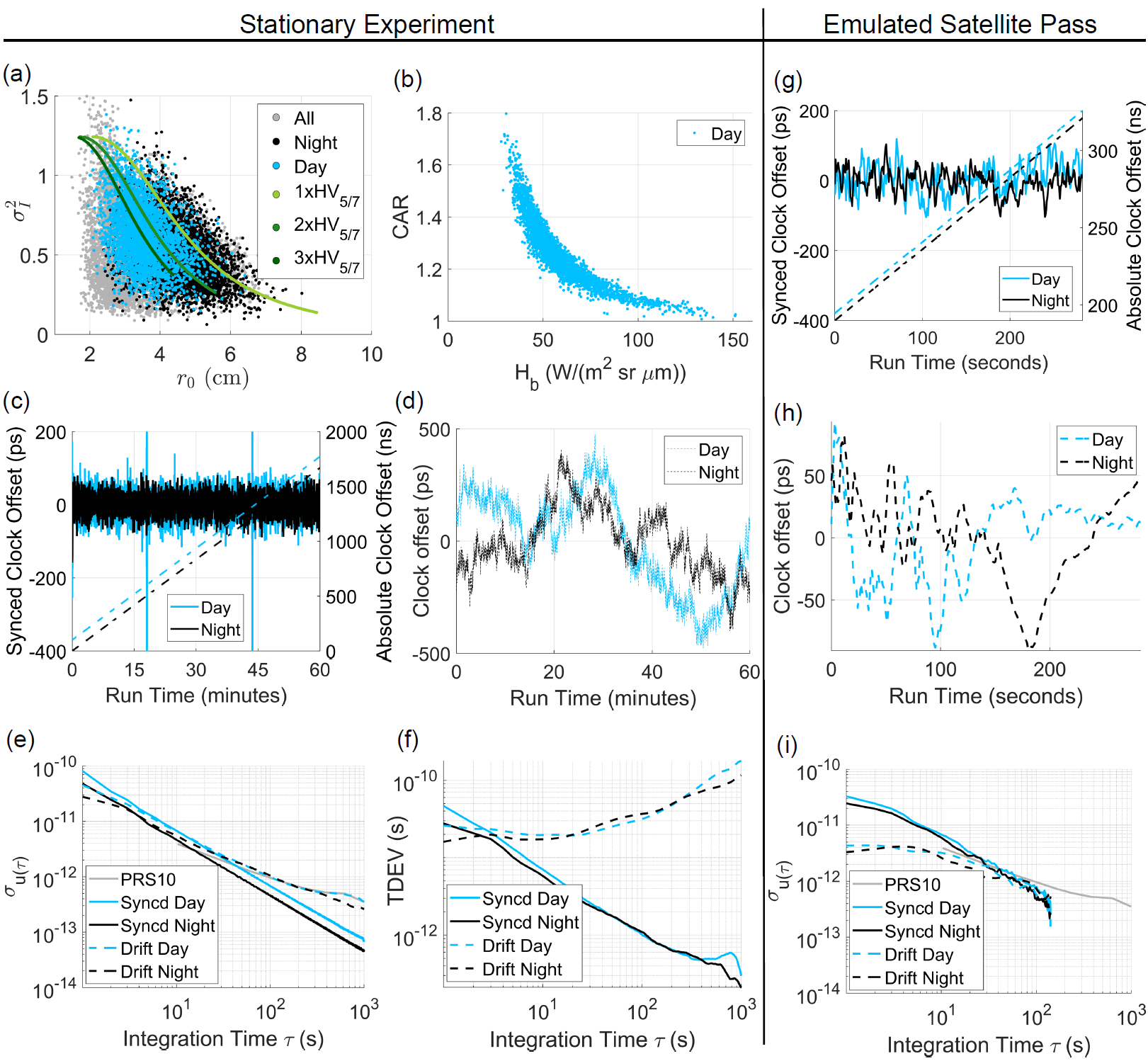}\\
	\caption{\label{fig:ONEPAGER}
	(a) The measured turbulence parameters [$r_0$, $\sigma_I^2$] and their projections onto the $1\times$, $2\times$, and $3 \times HV_{5/7}$ turbulence profiles \cite{gruneisen2021adaptive}.  
	The representative 1-hr acquisitions under daytime and nighttime conditions are in blue and black, respectively. 
	All other data is shown in gray. 
	(b) The CAR as a function of the background sky radiance $H_{\mathrm{b}}$ for the daytime scenario. 
	The (c) absolute offset of the synchronized clock (solid curves and left axis) and the drifting clock (dashed curves and right axis). 
	In (d) we take the drifting clock offset measurements and remove the linear trend to highlight the fluctuations in picoseconds. 
	In (e) and (f) we plot the overlapping Allan deviation and time deviation measured by the synchronized (solid) and drifting (dashed) clocks, respectively.  
	The gray line is the Allan deviation quoted in the PRS10 RbFS manual.
	In (g) -- (i) we present analogous plots for an emulated satellite pass. 	
	}
\end{figure*}

Figure~\ref{fig:schematic} gives a schematic of the two-way QTT components.
The sites labeled Alice and Bob each have a Thor Labs SPDC810 bi-photon source and a pair of Excelitas SPCM-AQRH detectors labeled $L$ and $R$ for local and receive, respectively, paired with 10-nm spectral filters.
The clock system is comprised of a Picoquant Hydraharp 400 time tagger, a Stanford Research Systems PRS10 Rubidium frequency standard (RbFS), and a PC.
Alice and Bob each randomly create a pair of photons, detect one of the pair with their local detector $L$, and send the other photon across the 1.6 km freespace channel to the other site where it is detected with a receive detector $R$.
Later we will emulate satellite motion in the field experiment data, so we write the detection times $t$ in the ``downlink'' direction $\alpha$ and ``uplink'' direction $\beta$ as
\begin{equation}\label{eq:RecieveTimes}
\begin{split}
t_{B,R} &= t_{A,L}+T_{\mathrm{prop}}^{(\alpha)}(t_{A,L}) + \delta\\
t_{A,R} &= t_{B,L}+T_{\mathrm{prop}}^{(\beta)}(t_{B,L}) - \delta,
\end{split}
\end{equation}
where the subscripts $A$ and $B$ correspond to the local clocks at Alice and Bob, respectively, $T_{\mathrm{prop}}^{(i)}$ are the directionaly dependant propagation times, and $\delta$ is the absolute clock offset where we make the approximation $\delta=\delta^{(\alpha)}=\delta^{(\beta)}$.
Rearranging Eq.~\ref{eq:RecieveTimes} one can define the relative clock offsets $\tau_{\alpha}$ and $\tau_{\beta}$: 
\begin{equation}
\begin{split}
\label{equ: rel clock offset}
\tau_\alpha &= t_{B,R}-t_{A,L} = T_{\mathrm{prop}}^{(\alpha)}(t_{A,L})+\delta \\
\tau_\beta &= t_{A,R}-t_{B,L} = T_{\mathrm{prop}}^{(\beta)}(t_{B,L})-\delta.
\end{split}
\end{equation}
Combining Eq.~\ref{equ: rel clock offset} we find the absolute clock offset \cite{lee2019symmetrical}
\begin{equation}
\begin{split}
\label{equ: abs clock}
\delta &= \dfrac{1}{2} \Big(\tau_\alpha-\tau_\beta-T_{\mathrm{prop}}^{(\alpha)}(t_{A,L})+T_{\mathrm{prop}}^{(\beta)}(t_{B,L}) \Big).
\end{split}
\end{equation}

In our field experiment we performed two-way QTT over a range of atmospheric conditions.
In Fig.~\ref{fig:ONEPAGER}(a) we show the measured turbulence parameters [$r_0$, $\sigma_I^2$], where each data point represents a one second average.
The projections onto the Hufnagel Valley $1\times$, $2\times$, and $3\times HV_{5/7}$ theoretical turbulence profiles \cite{gruneisen2021adaptive} show that the atmospheric conditions were similar or worse than an Earth-satellite downlink.
For further analysis, we select two continuous 1-hr data sets from the total data (gray) that are representative of daytime (blue) and nighttime (black) atmospheric conditions.
The average true coincidence rate for the two conditions and two directions is a modest $\sim$1000 cps.
Figure~\ref{fig:ONEPAGER}(b) shows the CAR of the QTT correlation signal in the $\alpha$ direction as a function of the background sky radiance $H_{\mathrm{b}}$ meausured during the daytime scenario; this indicates that the QTT algorithm reliably finds the true correlation signal with a bright sky background even as the CAR approaches 1. To account for Earth albedo and scattering \cite{coakley2003reflectance} in the uplink channel, we add noise to the receive detector data stream \cite{lafler2023quantum} corresponding to $\sim$25 W/(m$^2$$\,$sr$\,$$\mu$m).

For our initial analysis, we consider the timing performance of our two-way QTT clock system in the stationary field experiment.
We do this by utilizing the following two logical or ``software'' clocks.
One is drifting, that is, the frequency standards drift apart and the absolute clock offset $\delta$ is simply tracked.
For the synchronized clock we perform the following recursive algorithm where the acquisition time $T_a = 1$ s \cite{lafler2023quantum}: for the $i$'th acquisition we estimate the fractional-frequency drift $\Delta U^{(i)}_j = (\tau_j^{(i)}-\tau_j^{(i-1)})/T_{a}$ in each direction $j$, we shift the new received time tags according to $t_j^{(i+1)} + \Delta U^{(i)}_j\,T_a$, perform the QTT correlations, then use Eq.~\ref{equ: abs clock} to find the absolute offset $\delta^{(i+1)}$.

The results are shown in Fig.~\ref{fig:ONEPAGER}(c) for the daytime and nighttime scenarios with the synchronized (solid curves and left axis) and drifting (dashed curves and right axis) clocks.
The standard deviations are 27.1 and 39.7 ps for the nighttime and daytime scenarios, respectively.
As expected, the timing jitter is smaller at night.
Meanwhile, the drifting clock shows that the local clocks drift from each other at a rate of $\sim$450 ps per second on average.
Figure~\ref{fig:ONEPAGER}(d) shows the absolute clock offset $\delta$ according to the drifting clock after the mean clock drift is removed; this reveals the jitter of the QTT system, which is a consequence of the hardware components and atmospheric effects \cite{SupplementalMaterial}.
Altogether, we see that the two-way QTT algorithm can run continuously without losing synchronization for an extended period of time under challenging channel conditions.
\begin{figure}[t!]
	\includegraphics[width=0.85\columnwidth]{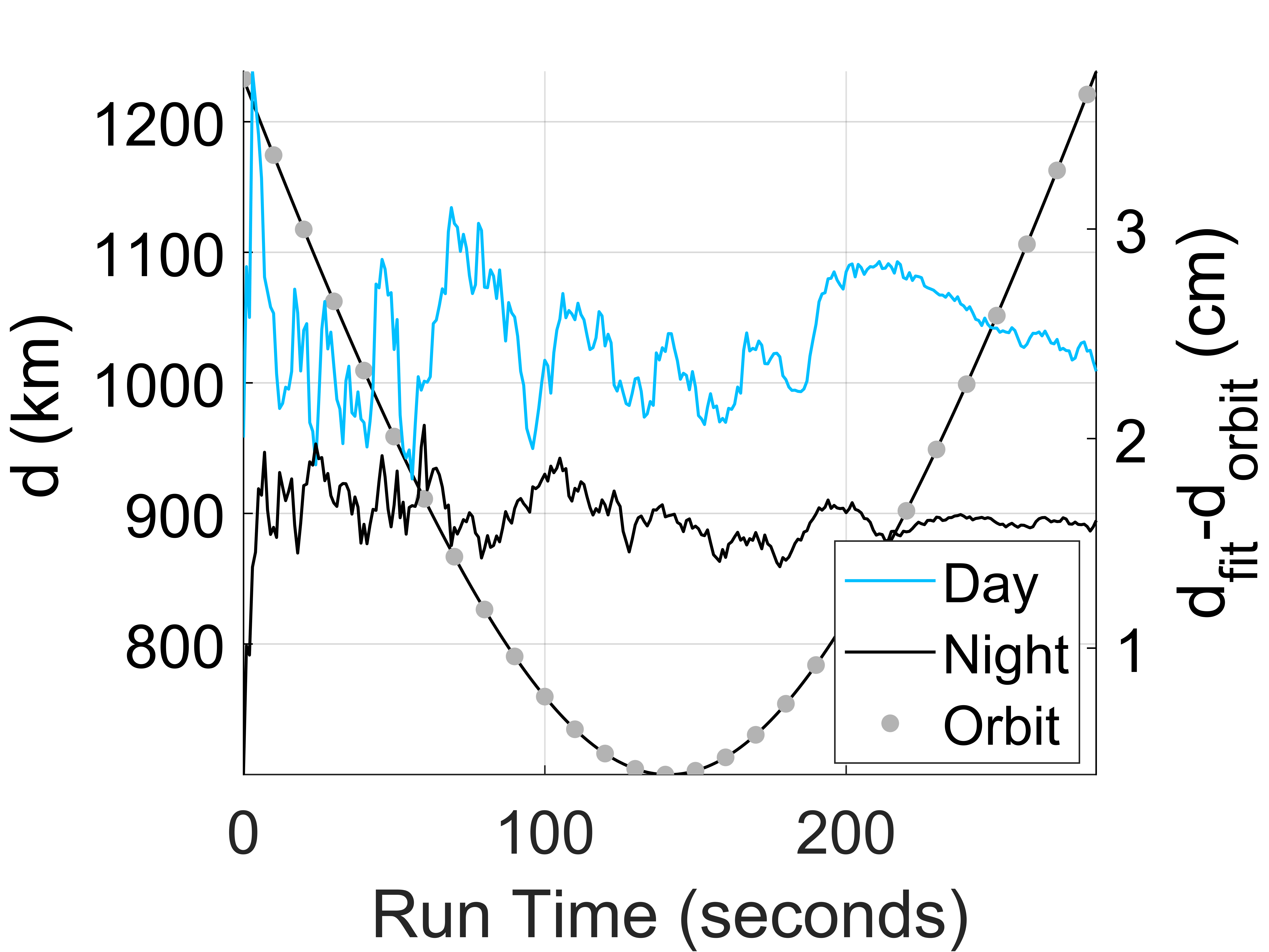}\\
	\caption{\label{fig: hist prop}
	Satellite range $d$ versus run time for the emulated satellite pass in the downlink direction. The left axis and gray points show the emulated range and the intersecting lines are the measured range for the daytime and nighttime conditions. The right axis is the difference between the measured range $d_{\mathrm{fit}}$ and the emulated range $d_{\mathrm{orbit}}$ showing cm-scale fluctuations.
	}
\end{figure}

The Allan deviation is a standard method to characterize the stability and noise profile of a clock system \cite{riley2008handbook}.
In Fig.~\ref{fig:ONEPAGER}(e) we show the overlapping Allan deviation, where the solid and dashed curves are the synchronized and drifting clocks, and the daytime and nighttime scenarios are blue and black, respectively.
The gray curve is the Allan deviation reported in the user manual of the Stanford Research Systems PRS10 RbFS.
The slopes for the drifting and synchronized clocks are approximately $-0.6$ and $-1$, respectively.
In Fig.~\ref{fig:ONEPAGER}(f) we show the time deviations (TDEV) \cite{riley2008handbook}, which have slopes approximately $+1/2$ and $-1/2$ for the drifting and synchronized clocks, respectively.
Combined these results indicate that the predominate noise is white FM and white PM for the drifting and synchronized clocks, respectively.
Visually, one can see these different noise profiles by comparing Figs.~\ref{fig:ONEPAGER}(c) and (d).

Since the Alice and Bob sites are stationary, we emulate relative motion by shifting the recieved time tags commensurate with a 700-km circular sun-synchronous orbit passing over head from -30 to 30 degree elevation \cite{SupplementalMaterial}.
This induces a large apparent frequency drift that broadens the QTT correlation signal \cite{lafler2023quantum}.
If the coincidence rate is large enough, one can simply reduce the acquisition time to narrow the correlation signal and perform the conventional QTT algorithm \cite{lafler2023quantum, haldar2023synch, haldar2023towards}.
However, for our proof-of-principle demonstration, the coincidence rates are not sufficiently high. 
Consequently, we create an orbit tracking algorithm by parameterizing the orbit equation into a fitting function and use an iterative process to find the orbit altitude $a$, inclination $\theta_i$, fractional frequency drift $\Delta U$, and clock offset $\delta$ \cite{SupplementalMaterial}. 
Figures~\ref{fig:ONEPAGER}(g)--(i) show the analogous plots for the emulated satellite pass. 
Removing the effect of relative motion recovers the QTT coincidence signal and constitutes the drifting logical clock (see Fig.~\ref{fig:ONEPAGER}(g) dashed line). 
Removing the clock drift $\Delta U$ as well constitutes the synchronized logical clock (see Fig.~\ref{fig:ONEPAGER}(g) solid line).
The standard deviations of the synchronized clock are 43.2 and 46.9 ps for the nighttime and daytime scenarios, respectively.
Figure~\ref{fig:ONEPAGER}(i) shows the noise profile as a consequence of the orbit tracking algorithm (for more details see Ref.~\cite{SupplementalMaterial}).
In Fig.~\ref{fig: hist prop} we plot the slant range for the entire pass in the downlink direction.
The gray points are the range of the emulated satellite orbit and the black (blue) lines are the measured range for the nighttime (daytime) conditions.
The left axis gives the range in kilometers and the right axis gives the difference between the measured range $d_{\mathrm{fit}}$ and the emulated range $d_{\mathrm{orbit}}$ in centimeters.

In this letter we report a remote clock synchronization and ranging field experiment over daytime space-Earth channel conditions utilizing a protocol we call two-way quantum time transfer.  
Our algorithm synchronizes to 10's of ps after only 1 second of integration with modest photon sources and detection equipment.
We analyze the performance of the protocol using the Allan and time deviations with synchronized and drifting software clocks.
We establish the relevance to space-Earth links using a software-emulated LEO satellite pass to $\mathcal{O}(1/c)$.
This work demonstrates a practical way to establish synchronization over a quantum link without unnecessarily complicating the system with extraneous time transfer methods.  
Follow on research should include smaller relativistic effects to $\mathcal{O}(1/c^3)$, further optimization of the orbit-tracking algorithm, and modelling of more complicated tasks such as entanglement swapping.

\section{Supplemental Material}
Relative motion is a challenge for any time transfer system. 
Keeping time between a ground station and a satellite in LEO is complicated by a large first-order Doppler shift $\mathcal{O}(1/c)$ and smaller relativistic effects \cite{blanchet2001relativistic}.
We studied how the QTT algorithm performs when the clock drift is small with respect to the QTT correlation signal \cite{lafler2023quantum}.
In this case, clock drifts $\Delta U \sim 10^{-10}$ do not significantly effect the QTT correlation signal (see Eq.~A7 in Ref.~\cite{lafler2023quantum}). 
Relativistic effects fall into this category and can be readily tracked by the standard QTT algorithm to picosecond level precision.
However, it may be useful to include them in the tracking algorithm discussed in Sec.~\ref{sec:RemSatMot}, but this is a topic of future research.
In the current analysis, it is important to show how the QTT algorithm performs with large first-order Doppler shifts that induce much larger clock drifts $\Delta U \sim 10^{-5}$.

In the 1.6-km testbed, traceability to daytime slant-path channels was achieved by rigorously characterizing and tuning the atmospheric turbulence and radiance conditions in the channel to match those of daytime slant-path propagation from space \cite{gruneisen2021adaptive}. 
We also introduced beam divergence to create 11 dB of aperture coupling loss representative of diffraction effects over
a 700-km propagation distance.
Thus, for this experiment, we emulate a 700-km circular sun-synchronous orbit with an inclination $\theta_i$ of 98.2 degrees.

\subsection{Emulating Satellite Motion} \label{sec: static}
\begin{figure}[t!]
	\includegraphics[width=1.0\columnwidth]{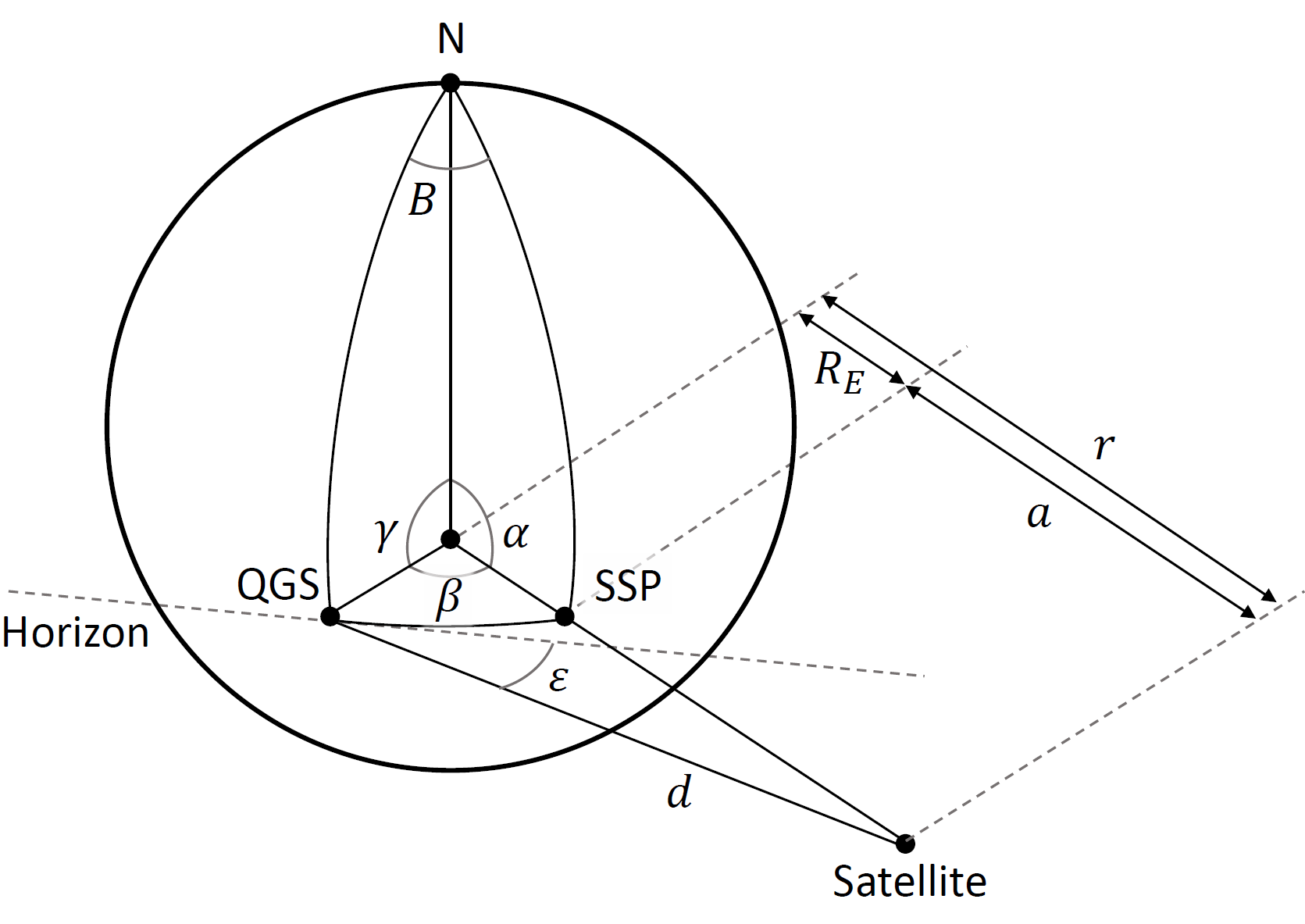}
	\caption{\label{fig:OrbitGraphic}
		Graphic relating the geocentric angle $\beta$ to the elevation angle $\varepsilon$ and the slant range $d$. On the surface of the Earth there is a spherical geometry relating $\beta$ to the central angles $\alpha$ and $\gamma$, which are functions of the latitudes $\lambda$, and the outer angle $B$, which is a function of the longitudes $\phi$ (see Eqs.~\ref{eq:inner} and \ref{eq:outer}). The two geometries are defined in terms of the north pole N, the location of the quantum ground station QGS, the satellite location, and the sub-satellite point SSP.
	}
\end{figure}

We introduce the first-order Doppler effect by shifting Alice's and Bob's receiver time tags according to the propagation time
\begin{equation} \label{equ: PropTime}
T_{\mathrm{prop}} = d/c,
\end{equation} 
where $c$ is the speed of light and the slant range $d$ is \cite{cakaj2011range}
\begin{equation} \label{equ: SlantRange}
d =  \sqrt{r^2 + R_{\mathrm{E}}^2 - 2 \, r R_{\mathrm{E}} \cos\beta} , 
\end{equation} 
$R_{\mathrm{E}}$ is the Earth's radius, $a$ is the altitude, $r \equiv R_{\mathrm{E}} + a $, and $\beta$ is the geocentric angle between the quantum ground station (QGS) and the satellite. 
From spherical trigonometry
\begin{equation}\label{eq:beta}
\begin{split}
\cos \beta &=   \cos \alpha \, \cos \gamma + \sin \alpha \, \sin \gamma \, \cos B,
\end{split}
\end{equation}
where the central angles are related to the latitudes $\lambda$
\begin{equation}\label{eq:inner}
\begin{split}
\alpha &= \pi/2 - \lambda_{\mathrm{SSP}}\\
\gamma &= \pi / 2 - \lambda_{\mathrm{QGS}},\\
\end{split}
\end{equation}  
the outer angle $B$ is related to the longitudes $\phi$
\begin{equation}\label{eq:outer}
B = \phi_{\mathrm{QGS}} - \phi_{\mathrm{SSP}},
\end{equation}
and SSP and QGS stand for sub-satellite point and quantum ground station, respectively (see Fig.~\ref{fig:OrbitGraphic}).

To find the slant range $d$ as a function of time we need the latitude and longitude of the satellite as functions of time.
Using Napier's rules for spherical right triangles (see Fig.~2) we have
\begin{equation}
\begin{split}
\sin  \lambda_{\mathrm{sat}}(t) &= \sin \omega_0 t \, \sin \theta_i\\
\tan  \phi_{\mathrm{sat}}(t) &= \tan \omega_0 t \, \cos \theta_i.
\end{split}
\end{equation}
Thus, for the latitude we find
\begin{equation}
\begin{split}
\lambda_{\mathrm{sat}}(t) &= \arcsin \big[ \sin (\omega_0 t + \Lambda_0)\sin \theta_i \big] \\
\Lambda_0 &= \arcsin\big[ \sin\lambda_{\mathrm{sat},0} /  \sin \theta_i  \big]\\
\omega_0 &= \sqrt{G\,M / r^3},
\end{split}
\end{equation}
where $\lambda_{\mathrm{sat},0}$ is the intial latitude, $\theta_i$ is the inclination angle, $\Lambda_0$ is a phase shift setting the intial satellite latitude when $t=0$, $G$ is the gravitational constant, and $M$ is the mass of the Earth.
The longitude is 
\begin{equation}
\begin{split}
\phi_{\mathrm{sat}}(t) &=  \arctan \big[  \tan (\omega_0t + \Lambda_0 )\, \cos\theta_i  \big] + \phi_{\mathrm{shift}} + \phi_P\\
\phi_{\mathrm{shift}} &= \phi_{\mathrm{sat},0} - \arctan\big[ \tan \Lambda_0 \, \cos \theta_i \big],\\
\phi_P &= (\omega_p-\omega_E) \, t,
\end{split}
\end{equation}
where $\phi_{\mathrm{sat},0}$ is the initial longitude, $\phi_{\mathrm{shift}}$ sets the intial longitude ($t=0$) by correcting for the longitudinal shift induced by the phase shift $\Lambda_0$, $\omega_E$ is the angular velocity of the Earth's rotation, $\omega_p$ is the sun-synchronous precession, and $\phi_P$ is a shift due to the Earth's rotation and the satellite orbit precession.
For our simulation, we set the intial satellite latitude and longitude to the ground station location so that the satellite is over head at $t=0$.

\begin{figure}[t!]
	\includegraphics[width=1.0\columnwidth]{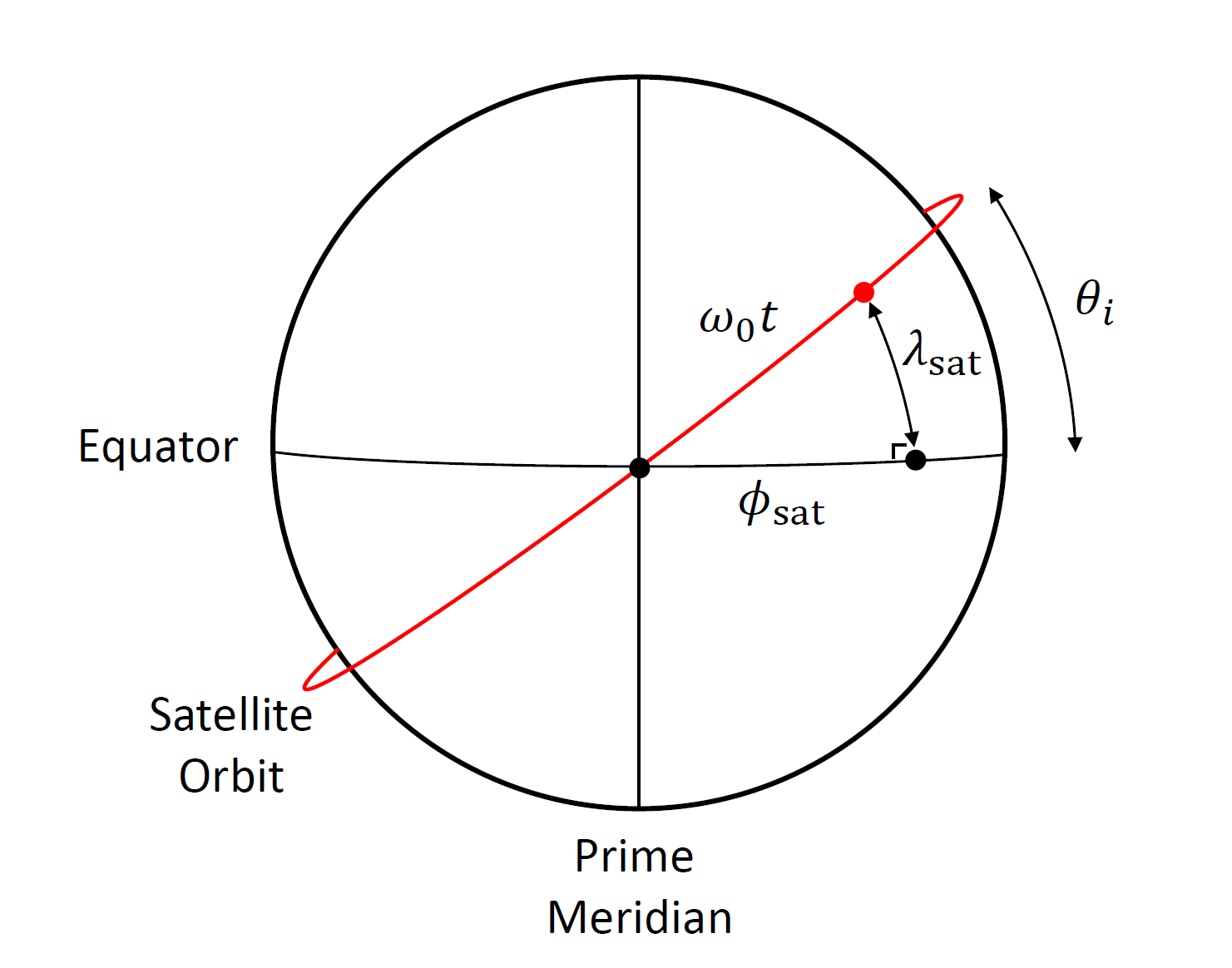}
	\caption{\label{fig:OrbitGraphic2}
		Graphic of spherical right triangle relating the latitude $\lambda_{\mathrm{sat}}$, longitude $\phi_{\mathrm{sat}}$, and geocentric angle $\omega_0 t$ for a circular orbit of inclination $\theta_i$.  
	}
\end{figure}

The uplink quantum signal must be pointed ahead of the downlink signal in order to intercept the satellite as it moves.
We use the point-ahead approximation
\begin{equation}\label{eq:PointAheadApprox}
\theta_{\mathrm{PA}} = \dfrac{2}{c} \, \bigg( \dfrac{G\,M}{r} \bigg)^{1/2} \, \sin \varepsilon.
\end{equation}
This expression can be written as a function of time by solving $d \sin \varepsilon = r \cos \beta - R_{\mathrm{E}}$ for $\varepsilon$ and substituting $\cos \beta$ from Eq.~\ref{eq:beta}. 
Consequently, the propagation time is asymmetric; $T_{\mathrm{prop}}$ is different in the uplink and downlink directions.

\subsection{Removing Satellite Motion}\label{sec:RemSatMot}
\subsubsection{Coarse Orbit Estimation}
\begin{figure}[t!]
	\includegraphics[width=1.0\columnwidth]{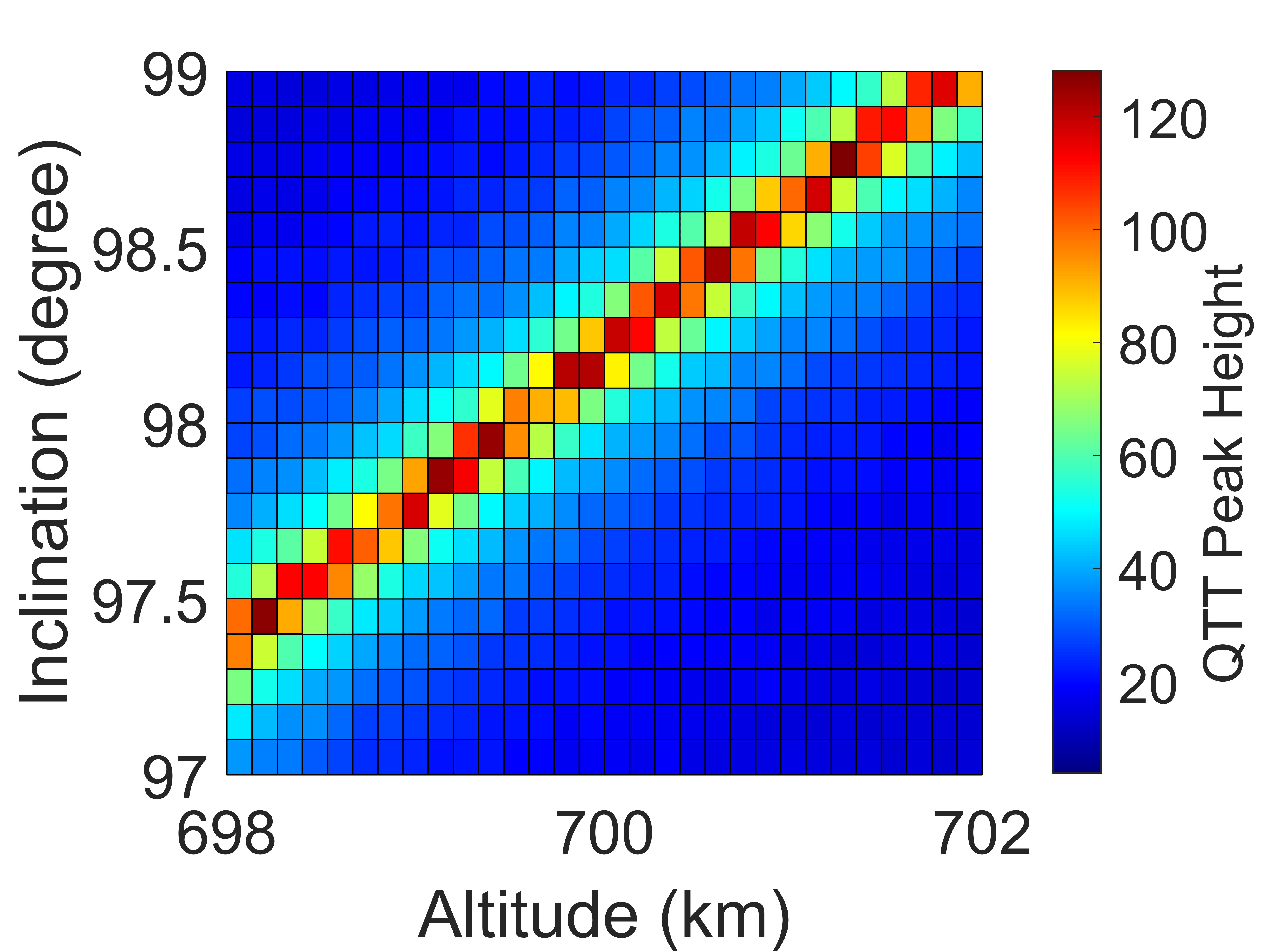}
	\caption{\label{fig:2D Scan}
		Coarse search over the orbit inclination and altitude parameters. The color legend gives the QTT correlation peak height. The step sizes are 0.1 degrees and 133 m, respectively.
	}
\end{figure}

For a LEO satellite, the rate of relative motion during our 1-second acquisition time prohibitively broadens the $\sim$1000 cps coincidence signal and makes it impossible to perform standard QTT \cite{lafler2023quantum} without pre-compensation.
Thus, the first step of our algorithm is to take the up and downlink directions and perform a coarse scan over altitude $a$ and inclination $\theta_i$ with a sufficiently large parameter space. 
For each $a$ and $\theta_i$, the corresponding $T_\mathrm{prop}(t,a,\theta_i)$ is calculated and subtracted from the received time tags, QTT is performed, and the correlation signal peak height is recorded.  
The result is an island of acceptable parameters where the correlation peak is large and the compensation is accurate enough to isolate the coincidences.
For example, Fig.~\ref{fig:2D Scan} shows the parameter scan for the nighttime-downlink senario.
Each red cell corresponds to an orbit-parmater pair $[a,\theta_i]$ that permits QTT to find the coincidences and move to the next step, despite the ambiguity with respect to the orbit.

\begin{figure*}[t!]
	\includegraphics[width=0.8\linewidth]{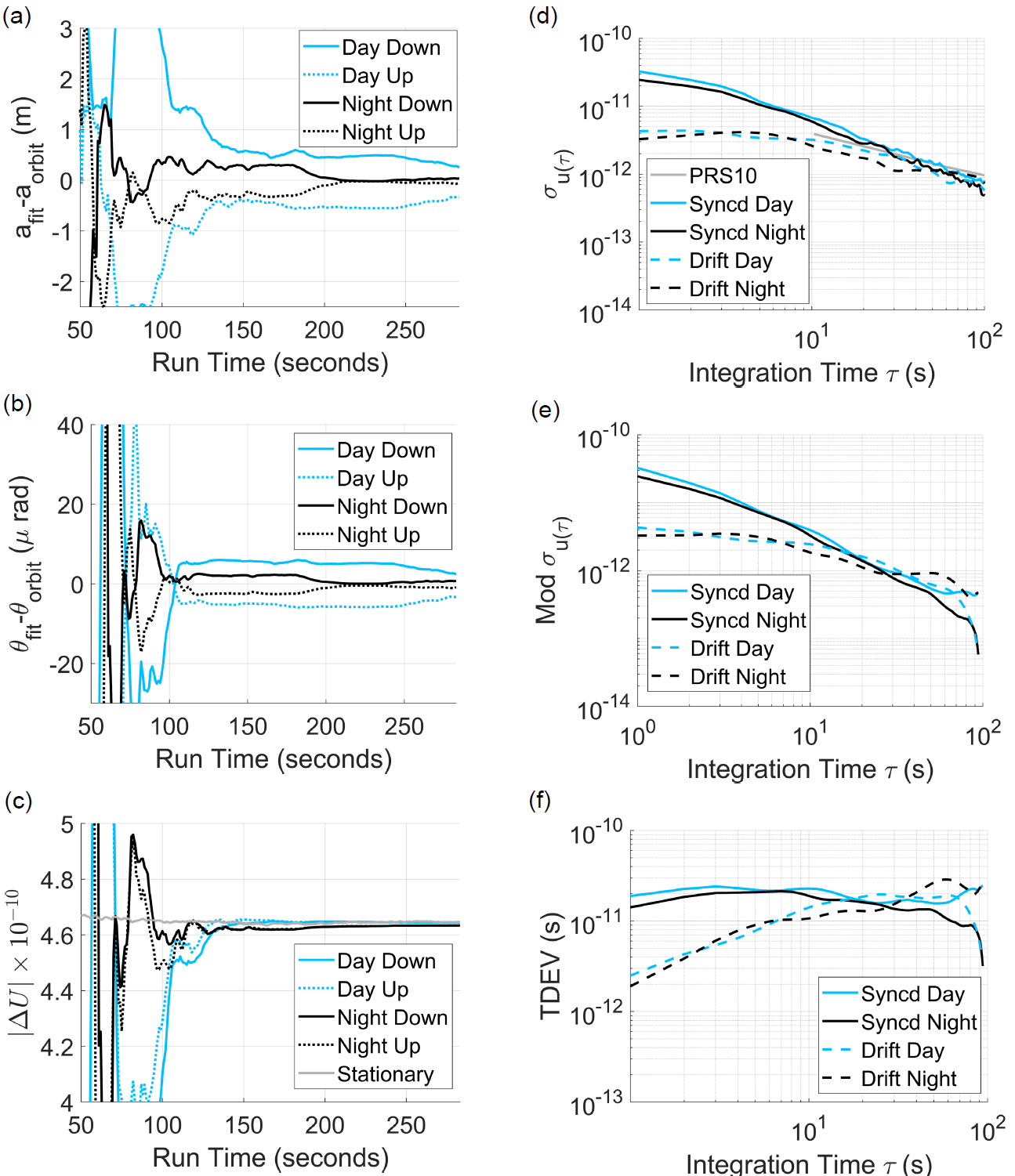}\\
	\caption{\label{fig:ONEPAGER}
		Plots summarizing the performance of the orbit tracking algorithm and QTT. (a)--(c) show how the fit algorithm converges to the altitude $a$, inclination $\theta_i$, and fractional frequency drift $\Delta U$, respectively. (d)--(e) give the Allen deviation, modified Allen deviation, and the time deviation, respectively.
	}
\end{figure*}	
\subsubsection{Precise Orbit Estimation}
Next, we take the coincidences found in the previous step and shift the received time tags back to their uncompensated locations.
We know that the time difference between coincident detection events is dominated by the Earth-satellite propagation time, but there is also the small linear clock drift inherent to the two PRS10 frequency standards.
Thus, we take the time difference between coincident time tags $\tau$ and fit them with 
\begin{equation}
\begin{split}
\tau_{\mathrm{fit}} \equiv T_{\mathrm{prop}}(t,a,\theta_i) +  m\,t + b,
\end{split}
\end{equation}
where the fit parameter $m$ accounts for the fractional frequency drift $\Delta U$ and $b$ accounts for the small vertical shift.
This fit results in a more precise measurement of the orbit parameters $a$ and $\theta_i$ as well as the frequency drift $\Delta U$; thus the coarse orbit determination only needs to be performed once at the beginning of the pass.
We perform the fit for the up- and down-link directions, where the uplink $T_\mathrm{prop}$ also includes the point ahead term in Eq.~\ref{eq:PointAheadApprox}.

\begin{figure*}[t!]
	\includegraphics[width=0.8\linewidth]{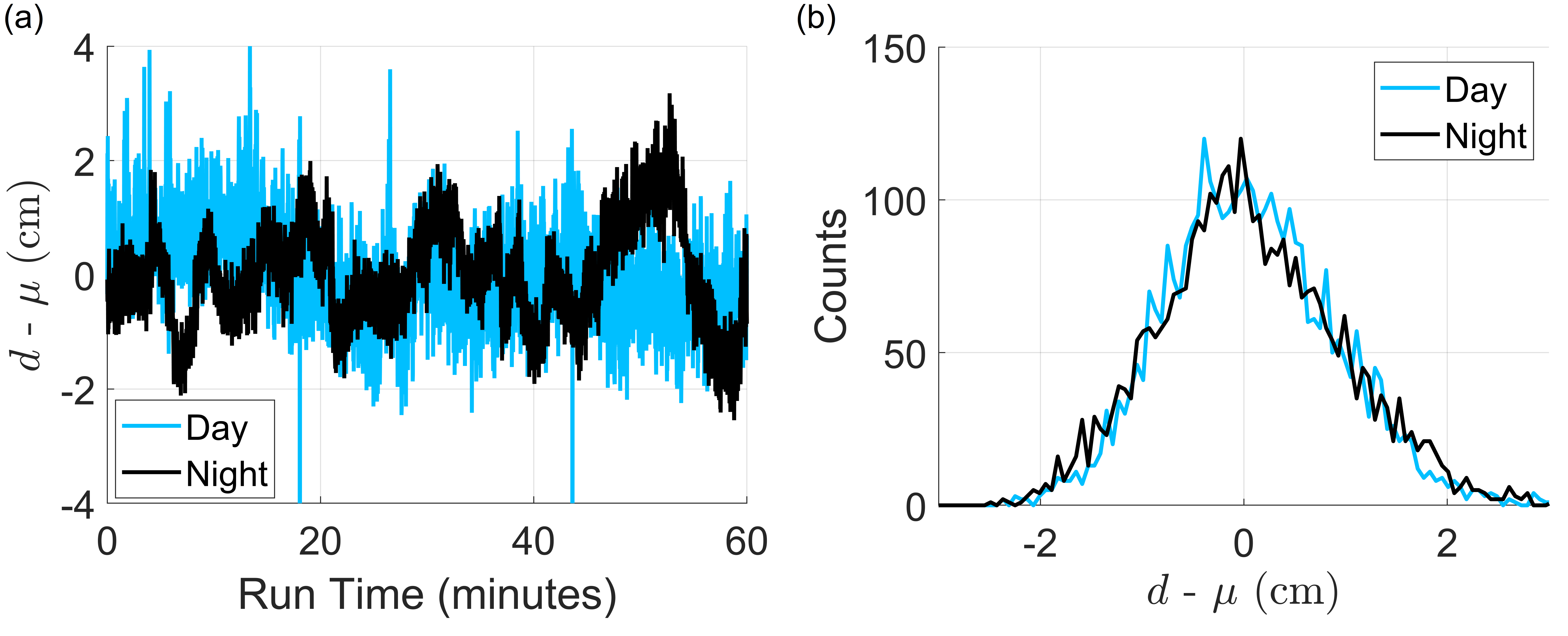}\\
	\caption{\label{fig: hist prop}
		Difference between the measured terrestrial testbed range $d$ and the mean range $\mu$=1.6445 km for the daytime (blue) and nighttime (black) scenarios. In (a) we give the range as a function of time and in (b) we histogram the one hour data acquisitions. The standard deviation is $\sim$0.9 cm for both scenarios.
	}
\end{figure*}	
\subsubsection{Software Clocks}	
The following recursive algorithm constitutes the synchronized clock: for each successive acquisition, the up- and down-link $\tau_\mathrm{fit}$ are used to remove the spreading of the time tags and align the clock, QTT is performed to measure the residual clock offset, isolate the new coincidences, add them to the prior coincidences, and update the fit.
The drifting clock follows from the same fit, except that only the orbit components are removed; thus the clocks drift apart according to the fractional frequency drift $\Delta U$.
In the next subsection we show how the algorithm converges to the input altitude, inclination, and the drift between the frequency standards.  

\subsubsection{Algorithm Performance}
In this subsection we summarize the performance of the orbit tracking algorithm and QTT.
Figures~\ref{fig:ONEPAGER} (a)--(c) show how the fit algorithm converges to the orbit altitude $a$, inclination $\theta_i$, and fractional frequency drift $\Delta U$, where the up- and downlink fit parameters are dashed and solid, respectively.
Figures~\ref{fig:ONEPAGER} (a) and (b) show that the fit parameters $a_\mathrm{fit}$ and $\theta_\mathrm{fit}$ slowly converge to the input parameters $a_\mathrm{orbit}$ and $\theta_\mathrm{orbit}$.
Figure~\ref{fig:ONEPAGER} (c) shows the up- and downlink linear drift parameters converging to the clock drift of the Rubidium frequency standards in about the same time.
The up- and downlink orbit-fit parameters are mirror images of each other due to the interplay with the fractional frequency drift $\Delta U$.
For the downlink direction, $\Delta U$ is positive and partially counteracting the effect of relative motion; this causes the fit to overestimate the altitude. 
The opposite is true for the uplink direction. 
Considering the nightime case in Figs.~\ref{fig:ONEPAGER} (a)--(c), one can see that after the satellite passes overhead and enough coincidences are collected, near $t$=200 s, the fitting algorithm can distinguish the two independent phenomenon and converges to the correct values.
The daytime case converges slower due to a noisier coincidence signal.  

Fig.~\ref{fig:ONEPAGER} (d)--(e) give the Allen deviation, modified Allen deviation, and the time deviation, respectively.
Comparing these results to the corresponding figures in the main text, one can see that the synchronized clocks perform similarly to the drifting clocks in the stationary senario. The slope of the modified Allan deviation (Fig.~\ref{fig:ONEPAGER} (e)) is $\approx -1$, which indicates flicker PM  noise, whereas the stationary synchronized clocks have white PM noise \cite{riley2008handbook}.
We suspect that the noise profiles are different due to the time it takes for the fit parameters to converge to the emulated orbit parameters.
In other words, if the tracking algorithm was further optimized or if the satellite pass were longer, we suspect the slope of the modified Allen deviation would change to -1.5 and indicate white PM noise.
This also explains why the drifting clocks are worse than the stationary senario.

\subsubsection{System Jitter}
%In Fig.~\ref{fig: hist prop} we plot the range of the terrestrial testbed field experiment for the nightime and daytime 1-hour acquisitions.
%We use $d = c \times T_{\mathrm{prop}}$ where $T_{\mathrm{prop}}=(\tau_{\alpha}+\tau_{\beta})/2$ is calculated using the relative offsets $\tau$ from the drifting clock.
%If we were to use the synchronized clock relative offsets, then the system jitter would be partially corrected for when $\Delta U$ is measured and applied in the predict-ahead algorithm.
%We find that the mean of both distributions is $\mu = 1.6445$ km with standard deviation $\sim$0.9 cm (30 ps) for the daytime and nighttime scenarios.
%The distribution in Fig.~\ref{fig: hist prop}(b) is quite narrow because the noise from the frequency standards, subsumed in $\delta$, cancels due to the symmetry of the two-way protocol. 
%Thus, the jitter observed in Fig.~\ref{fig: hist prop} is a consequence of the atmosphere and detection equipment. 
%We subtract the mean propagation distance from the data before emulating the satellite pass, but the system jitter is not corrected for.
%Thus, all the jitter of the terrestrial field experiment, including the frequency standard jitter, is present in the emulated satellite-pass.
In Fig.~\ref{fig: hist prop} we plot the range of the terrestrial testbed field experiment for the nightime and daytime 1-hour acquisitions.
We use $d = c \times T_{\mathrm{prop}}$ where $T_{\mathrm{prop}}=(\tau_{\alpha}+\tau_{\beta})/2$ is calculated using the relative offsets $\tau$ from the drifting clock.
%If we were to use the synchronized clock relative offsets, then the system jitter would be partially corrected for when $\Delta U$ is measured and applied in the predict-ahead algorithm.
We find that the mean of both distributions is $\mu = 1.6445$ km with standard deviation $\sim$0.9 cm (30 ps) for the daytime and nighttime scenarios.
The observed jitter is mainly due to the atmosphere and the detection equipment since the noise from the frequency standards tends to cancel due to the symmetry of the two-way protocol.
On the other hand, when $T_{\mathrm{prop}}$ is calculated with respect to the synchronized clock, the noise is mostly white noise, which suggests that the system jitter is partially corrected for when $\Delta U$ is measured and applied in the predict-ahead algorithm.
Therefore, the later method could be used to enhance the measurement of $T_{\mathrm{prop}}$ and thus the range $d$.
Nevertheless, we simply take the data and subtract the mean propagation time observed in the test bed before emulating motion.
Therefore, all the jitter of the terrestrial field experiment, including the frequency standard jitter, is present in the emulated satellite-pass.

%\section{Acknowledgements}
%The views expressed are those of the author and do not necessarily reflect the official policy or position of the %Department of the Air Force, the Department of Defense, or the U.S. government.
%The appearance of external hyperlinks does not constitute endorsement by the U.S. Department of Defense (DoD) of the linked websites, or the information, products, or services contained therein. The DoD does not exercise any editorial, security, or other control over the information you may find at these locations. Approved for public release; distribution is unlimited. Public Affairs release approval AFRL-2024-0986.

\section*{Acknowledgements}
The authors acknowledge program management support from Valerie Knight, Jacob DeLange, Ryan Riley, and Ian Blake AFRL.
The views expressed are those of the author and do not necessarily reflect the official policy or position of the Department of the Air Force, the Department of Defense, or the U.S. government.
The appearance of external hyperlinks does not constitute endorsement by the U.S. Department of Defense (DoD) of the linked websites, or the information, products, or services contained therein. The DoD does not exercise any editorial, security, or other control over the information you may find at these locations. Approved for public release; distribution is unlimited. Public Affairs release approval AFRL-2024-0986.

%Bibliography
\bibliography{2_way_quantum_time_transfer_field_exp_v_arXiv_2}

\end{document}